\documentclass[preprint, 12pt,authoryear]{elsarticle}
\usepackage{natbib}

\usepackage[utf8]{inputenc}
\usepackage[T1]{fontenc}

\usepackage{amssymb}
\usepackage{amsmath}

\usepackage{float}
\makeatletter
\renewcommand{\fps@table}{H} 
\makeatother

\usepackage{threeparttable}
\usepackage{graphicx}%
\usepackage{multirow}%
\usepackage{textcomp}%
\usepackage{booktabs}%
\usepackage{hyperref}
\usepackage{geometry}
\usepackage{float}

\begin{document}

\begin{frontmatter}




\title{Rethinking AI Literacy Education in Higher Education: Bridging Risk Perception and Responsible Adoption} 
\tnotetext[tn1]{This is the accepted manuscript of the article accepted for publication in \textit{Social Sciences \& Humanities Open}. The final authenticated version will be available via the journal website.}
\author[inst1,inst2]{Shasha Yu\corref{cor1}} 
\author[inst2]{Fiona Carroll} 
\author[inst2,inst3]{Barry L. Bentley} 

\cortext[cor1]{Corresponding author}

\affiliation[inst1]{organization={School of Professional Studies, Clark University},
            addressline={950 Main Street}, 
            city={Worcester},
            postcode={01610}, 
            state={MA},
            country={USA}}

\affiliation[inst2]{organization={Cardiff School of Technologies, Cardiff Metropolitan University},
            addressline={Llandaff Campus, Western Avenue}, 
            city={Cardiff},
            postcode={CF5 2YB}, 
            state={Wales},
            country={UK}}
\affiliation[inst3]{organization={Harvard Medical School},
            addressline={25 Shattuck Street}, 
            city={Boston},
            postcode={02115}, 
            state={MA},
            country={USA}}




\begin{abstract}
As AI becomes increasingly embedded across societal domains, understanding how future AI practitioners---particularly technology students---perceive its risks is essential for responsible development and adoption. This study analyzed responses from 139 students in Computer Science, Data Science/Data Analytics, and other disciplines using both explicit AI risk ratings and scenario-based assessments of risk and adoption willingness. Four key findings emerged: (1) Students expressed substantially higher concern for concrete, explicitly stated risks than for abstract or scenario-embedded risks; (2) Perceived risk and willingness to adopt AI demonstrated a clear inverse relationship; (3) Although technical education narrowed gender differences in risk awareness, male students reported higher adoption willingness; and (4) A form of ``risk underappreciation'' was observed, wherein students in AI-related specializations showed both elevated explicit risk awareness and higher willingness to adopt AI, despite lower recognition of risks in applied scenarios. These findings underscore the need for differentiated AI literacy strategies that bridge the gap between awareness and responsible adoption and offer valuable insights for educators, policymakers, industry leaders, and academic institutions aiming to cultivate ethically informed and socially responsible AI practitioners.
\end{abstract}




\end{frontmatter}



\section{Introduction}\label{sec1}

Artificial Intelligence (AI) is reshaping modern society across sectors such as healthcare, education, transportation, and entertainment \citep{shaheen2021applications,zhai2021review}. Its integration delivers substantial benefits---including efficiency gains, new discoveries, and improved convenience \citep{lee2023benefits}---while simultaneously introducing well-documented risks such as privacy breaches, algorithmic bias, job displacement, and misuse \citep{yu2023balance}. Despite growing public exposure to AI, awareness of these risks remains uneven \citep{yu2024insights,brauner2023does}, at times leading to uninformed adoption and inadequate scrutiny \citep{yu2024chatgpt}. The rapid expansion of generative AI in both educational and professional settings has intensified the need to understand how emerging practitioners assess AI-related risks.

Within computing and engineering education, scholars emphasize that future AI practitioners require explicit preparation for ethical reasoning and responsible design practices \citep{kasinidou2021educating,brown2024teaching,archambault2024ethical}. Technology students—who will become the developers, analysts, and engineers responsible for building and governing AI systems—represent a particularly consequential population. Their perceptions of AI risks shape not only their own adoption behaviors but also the decisions they will make as future professionals. However, although prior studies have examined student attitudes toward AI \citep{saez2025students,STOHR2024100259} and perceptions of AI's benefits and risks in educational contexts \citep{li2025embracing,tierney2025student}, comparatively little work has jointly examined (a) explicit risk awareness, (b) scenario-based recognition of risks in applied contexts, and (c) how these relate to willingness to adopt AI technologies. Key questions therefore persist:

\begin{itemize}
\item What types of AI risks are tech students most concerned with? 
\item Are there patterns or gaps in their understanding of AI risks that need to be addressed?
\item How does their awareness of risks shape their willingness to adopt, develop, or advocate for AI technologies? 
\item Are there differences in the understanding of AI risks based on technical specialization?
\end{itemize}

Existing studies examine students' attitudes toward AI adoption \citep{saez2025students} and their perceptions of AI's benefits and risks in educational contexts \citep{li2025embracing,tierney2025student}. Some works investigate general student populations \citep{delello2023exploring} or center on adoption determinants rather than risk awareness \citep{wang2021factors}, leaving technology students' AI risk understanding comparatively underexamined. Research in computing education and FATE further documents uneven recognition of fairness and safety issues in real-world systems \citep{kasinidou2021educating,kasinidou2021agree,brown2024teaching,archambault2024ethical,pierson2017demographics}, underscoring the need for empirical work that situates risk awareness as a core construct in understanding responsible AI adoption.

This gap limits the development of targeted AI literacy programs that prepare students to navigate and mitigate risks responsibly. To address this need, the present study investigates the current state of AI risk awareness among technology students, examines patterns in explicit and scenario-based assessments, and analyzes how these perceptions shape willingness to adopt AI technologies. By integrating both explicit and scenario-based assessments, the study also provides a more comprehensive view of how students articulate risks in the abstract and how they recognize them in applied contexts.

By uncovering these relationships, this work seeks to inform curriculum design and policy development aimed at fostering responsible AI use and development. Ultimately, the study contributes to bridging the gap between rapid AI innovation and informed, ethical practice.

\section{Background}
\subsection{AI Risks and Challenges}\label{subsec:ai_risks}
A growing body of research highlights the multifaceted risks associated with AI systems across technical, ethical, and societal dimensions. Concerns about privacy and data security are among the most frequently discussed, particularly as AI technologies increasingly rely on large-scale personal data and often enable secondary uses beyond individuals' awareness \citep{yu2023enhancing,manikonda2018s,naude2020artificial}. Alongside privacy issues, scholars have extensively documented how algorithmic systems reproduce or even amplify social biases, leading to unfair outcomes in high-stakes settings such as hiring, lending, or policing \citep{roselli2019managing,raghavan2020mitigating,albaroudi2024comprehensive}. Technical vulnerabilities also represent an important class of risks: adversarial attacks and other forms of system manipulation can compromise the reliability of AI models deployed in critical domains \citep{chen2019adversarial,manikonda2018s}.

Beyond these technical concerns, AI's broader societal implications have also drawn increasing scholarly attention. Automation-related job displacement continues to raise questions about economic inequality and the need for large-scale workforce adaptation \citep{rawashdeh2025consequences,yu2023balance}. The rapid spread of misinformation—including deepfakes and synthetic media—illustrates how AI can be leveraged to distort public discourse and manipulate public opinion \citep{aimeur2023fake,al2021using,agarwal2020detecting}. In parallel, the expansion of AI-enabled autonomous weapons has intensified ethical and accountability debates, especially regarding the delegation of lethal decision-making to non-human agents \citep{de2020ethics,yu2022implications}. 

Lack of transparency further complicates responsible AI deployment. ``Black-box'' systems limit users' and regulators' ability to understand or contest automated decisions, raising particular concerns in areas such as criminal justice and healthcare \citep{larsson2020transparency,von2021transparency,bernal2022transparency}. These transparency challenges intersect with a broader set of ethical issues—including autonomy, consent, and civil liberties—that scholars have argued are central to assessing the real-world impact of AI technologies \citep{stahl2021ethical,benzinger2023should,saheb2023ethically}. AI systems also pose psychological and behavioral risks, influencing creativity, critical thinking, social interaction, and exposure to diverse perspectives through personalization and algorithmic filtering \citep{torres2024systematic,zhai2024effects,yu2024trust,park2024filter}. 

Finally, researchers have emphasized that AI has the potential to exacerbate existing social inequalities by embedding structural disadvantages present in training data or institutional contexts \citep{zajko2022artificial,yu2022insights,capraro2024impact}. These concerns intersect with the broader risk of AI misuse—ranging from targeted disinformation campaigns to surveillance and political manipulation—enabled by the scale and efficiency of contemporary AI systems \citep{anderljung2024protecting,yu2022implications}.  

Together, these strands of scholarship identify a comprehensive set of AI-related risks that inform the conceptual framing of this study and underpin the explicit AI risk awareness items used in the survey instrument.

\subsection{AI Risk Perception and Adoption}

Risk perception often diverges between abstract judgments and applied, context-dependent decisions. Foundational work shows that framing, salience, and bounded rationality shape how individuals weigh risks and benefits \citep{slovic2016perception,kahneman2013prospect}. Technology-adoption research similarly models how perceived risk moderates intention \citep{venkatesh2003user,ajzen1991theory}. In education settings, studies report mixed patterns in students' AI risk judgments and adoption tendencies, with gaps between perceived usefulness and concerns about accuracy, bias, and learning effects \citep{crockett2020risk,oc2024generative,schei2024perceptions}. These literatures motivate our use of both explicit and scenario-based measures: the former elicits articulated concerns about named risks; the latter probes applied recognition when risks are not made salient a priori, which prior research suggests can surface different judgments \citep{slovic2016perception,kahneman2013prospect}.

\subsection{AI Literacy Education in Higher Education}

Universities increasingly incorporate ethical and societal dimensions into computing curricula, yet implementations vary and often emphasize principles over applied risk judgment \citep{brown2024teaching}. Empirical studies show that while students can articulate aspects of fairness, accountability, and transparency, recognition of risks in realistic systems is uneven and sensitive to disciplinary identity and context \citep{kasinidou2021educating,kasinidou2021agree,pierson2017demographics,archambault2024ethical}. Policy frameworks likewise foreground responsible AI and critical AI literacy \citep{UNESCOAI2021,EUAIAct2024}, but evidence-based guidance on integrating scenario-based risk recognition into technical programs remains limited. This background motivates curricula that combine explicit instruction with applied, scenario-centered activities to strengthen transfer from concepts to practice \citep{kasinidou2021educating, archambault2024ethical}.

\subsection{AI Risk: The Missing Piece of a Tech Curriculum}
Although AI perceptions and adoption have been widely studied, most existing work examines the general public or broad student populations rather than technology students \citep{crockett2020risk,saez2025students,tierney2025student}. Prior studies often highlight general attitudes toward AI and concerns such as rapid technological change or unemployment, but they rarely assess concrete risk domains such as bias, privacy, or misuse in technically trained groups \citep{jeffrey2020understanding,ghotbi2021moral}. Research on ChatGPT adoption focuses on usability and credibility while overlooking students' awareness of potential risks \citep{masa2024antecedents}. As a result, relatively less is known about how different dimensions of AI risk awareness relate to students' willingness to adopt or develop AI systems.

This gap is important because technology students are the future developers, engineers, and decision-makers who will influence the ethical direction of AI \citep{yu2024insights}. Understanding their risk awareness is essential for clarifying how such awareness shapes responsible adoption and for determining whether students are prepared to incorporate ethical considerations into system design. Addressing this gap requires empirical study and educational models that explicitly connect risk perception, adoption behavior, and ethical development practices \citep{mittelstadt2019principles,floridi2010cambridge}.

This study responds to this need by examining technology students' awareness of a range of AI risks and their willingness to adopt AI technologies. The findings are intended to support AI ethics education and curriculum design that better prepare students for responsible AI development.

\section{Methodology}
\subsection{Research Design}
This study utilized a cross-sectional, mixed-methods survey to investigate technology students' awareness of AI risks and their willingness to adopt AI technologies. The instrument contained four sections: demographic characteristics, explicit AI risk awareness (quantitative), scenario-based evaluation (quantitative), and a short set of open-ended questions (qualitative) designed to elicit students' broader views on AI, perceived risks, and expectations for AI risk education. The full survey instrument is provided in the Supplementary Material.

The present manuscript analyzes only the quantitative components, which provide structured measures of explicit and scenario-based risk awareness and adoption willingness. The qualitative responses will be examined and reported in a separate manuscript, where they will offer complementary insight into students' interpretations of AI risks and their perspectives on educational needs.

\subsection{Sample and Participants}
This study targeted students and recent graduates (within one year) enrolled in AI/ML-related graduate programs (e.g., Computer Science, Data Science, Data Analytics) at a private university in the United States. These programs include structured coursework in artificial intelligence and machine learning and therefore provide access to students with substantial exposure to AI technologies.

Participants were recruited through multiple channels: (1) distribution of the anonymous Qualtrics survey link during class sessions with instructor permission, (2) the university's SONA Research Participation System, which allowed eligible volunteers to complete the study for course credit, (3) direct email invitations sent to students in the target programs, and (4) voluntary snowball sampling in which participants shared the anonymous survey link with peers. All data were collected online, and no referral incentives were offered.

A total of 189 responses were recorded between November 19, 2024, and February 19, 2025. After removing 50 incomplete or inconsistent submissions, and with no exclusions based on demographic characteristics, 139 valid responses remained. Of these, 116 participants were enrolled in AI-focused programs (Computer Science: $n=48$; Data Science/Data Analytics: $n=68$). An additional 23 respondents were obtained through snowball sampling from programs with varying degrees of AI exposure, including Information Technology, Human-Computer Interaction, Interactive Media, Game Design, Psychology, and Biology. These comparative groups were included to examine how differences in formal AI training relate to risk perception and adoption patterns.

Throughout this manuscript, ``CS'' refers to Computer Science, ``DS/DA'' to Data Science or Data Analytics, and ``Other'' to interdisciplinary or non-technical fields represented in the sample.

\textbf{Demographic profile.}  
The sample was predominantly young, with 63.3\% aged 18--24 and 34.5\% aged 25--34, and 2.2\% aged 35--44. Gender distribution was nearly balanced, with 51.1\% identifying as female and 48.9\% as male. The majority identified as Asian (77.7\%), followed by White or Caucasian (10.8\%), Hispanic or Latino (4.3\%), Black or African American (2.2\%), Mixed or Multiple Ethnicities (2.9\%), and 2.2\% who preferred not to disclose. Most participants were current graduate students (Master's or PhD, 75.5\%), with undergraduates comprising 21.6\%; recent graduates and working professionals formed a small minority. By specialization, 48.9\% reported Data Science/Data Analytics, 34.5\% Computer Science, and 16.5\% other fields. Regarding experience, 47.5\% reported 1--3 years in the technology field, 28.1\% less than one year, 21.6\% 4--6 years, and 2.9\% more than 7 years. Self-rated AI knowledge was generally high: 46.0\% moderate, 19.4\% proficient, 6.5\% expert, 24.5\% basic, and 3.6\% no understanding. See Table~\ref{tab:demographics} for full details.

\subsection{Materials and Measures}

The survey instrument consisted of four sections: 
(a) seven demographic items; 
(b) one matrix-style item assessing explicit awareness across twelve AI risk domains; 
(c) ten scenario-based items, each measuring both perceived risk and willingness to adopt the technology; 
and (d) six open-ended qualitative questions, which are analyzed in a separate manuscript. 
One attention-check item was included to ensure response quality.

\textbf{Explicit AI Risk Awareness.}  
Participants rated their concern for 12 explicitly named AI risk domains (1 = Not concerned at all, 5 = Very concerned). These domains reflect widely documented categories of AI-related risks across AI ethics, governance, and HCI scholarship. 

\textbf{Scenario-Based Risk Awareness and Adoption Willingness.}  
Ten realistic AI applications were presented without naming the underlying risks. For each application, participants rated (a) perceived risk and (b) willingness to adopt the system (1 = Not at all, 5 = Very). This design draws on research showing that risk judgments differ when individuals evaluate concrete contexts rather than abstract descriptions \citep{slovic2016perception,kahneman2013prospect}. Applications included hiring platforms, healthcare diagnostics, financial advisory tools, public surveillance, manufacturing automation, personalized learning, and smart-home systems.

Table~\ref{tab:risk_domains} provides an overview of how the twelve explicit AI risk domains correspond to the ten scenario-based applications. This mapping clarifies the conceptual alignment between explicit risk categories and their applied manifestations.

\begin{table}[H]
\centering
\small
\setlength{\tabcolsep}{6pt}
\caption{Mapping of Explicit AI Risk Domains to Scenario-Based Survey Items}
\label{tab:risk_domains}
\begin{tabular}{p{0.32\linewidth} p{0.60\linewidth}}
\toprule
\textbf{Risk domain} & \textbf{Scenario example (Survey item)} \\
\midrule
General AI risks & No direct scenario counterpart \\
Privacy / Data security & Smart-home assistant learning user habits (Q13) \\
Bias / Fairness & Job-matching platform ranking applicants (Q14) \\
Security vulnerabilities & AI financial advisory system analyzing spending patterns (Q15) \\
Job displacement & AI-driven automation in manufacturing (Q16) \\
Misinformation / Manipulation & AI-curated personalized news feeds (Q17) \\
Autonomous weapons & Autonomous drones used for automated threat response (Q18) \\
Transparency issues & AI diagnostic tool predicting medical conditions (Q19) \\
Ethical implications & AI-enabled public surveillance systems (Q20) \\
Psychological / Cognitive impacts & AI assistant managing tasks and routines (Q21) \\
Social inequalities & Adaptive learning platform personalizing content (Q22) \\
Misuse / Abuse & No direct scenario counterpart \\
\bottomrule
\end{tabular}

\vspace{6pt}
\begin{flushleft}
\footnotesize
\textit{Note.}  
Scenario descriptions correspond to Survey Items Q13--Q22 in Supplementary Material.  
Supporting literature for each explicit risk domain:  
privacy/data security \citep{yu2023enhancing,manikonda2018s,naude2020artificial};  
bias/fairness \citep{raghavan2020mitigating,roselli2019managing,albaroudi2024comprehensive};  
security vulnerabilities \citep{chen2019adversarial,manikonda2018s};  
job displacement \citep{rawashdeh2025consequences,yu2023balance};  
misinformation \citep{aimeur2023fake,agarwal2020detecting,al2021using};  
autonomous weapons \citep{de2020ethics,yu2022implications};  
transparency \citep{larsson2020transparency,von2021transparency};  
ethical implications \citep{stahl2021ethical,benzinger2023should,saheb2023ethically};  
psychological/cognitive impacts \citep{zhai2024effects,torres2024systematic,yu2024trust,park2024filter};  
social inequalities \citep{zajko2022artificial,yu2022insights,capraro2024impact}.  
General AI risks and Misuse/Abuse appear only as explicit items because they represent
broad, cross-cutting concerns that do not map cleanly to a single application context.  
\end{flushleft}
\normalsize
\end{table}

\subsection{Procedure}

The survey was administered online via Qualtrics, ensuring anonymity and data security. Participants provided informed consent through an electronic page detailing the study's purpose, voluntary nature, and confidentiality. Those declining consent were redirected to a closing page. In-class participants could opt for an alternative task, and SONA participants received course credits. A quality check question identified inattentive responses. 

This study was approved by the Institutional Review Board at Clark University (Protocol No. 701). All participants provided informed consent prior to participation. No personally identifiable information was collected.

\subsection{Validation}

The instrument underwent a two-stage validation process prior to deployment. First, three faculty members with expertise in AI ethics and human-computer interaction reviewed all items for clarity, relevance, and completeness. Their feedback informed revisions to item wording and alignment across risk domains. Second, a small cognitive pilot with 12 technology students was conducted to evaluate item comprehension, survey flow, and response burden. Minor adjustments were made based on participants' comments before releasing the final instrument.

\subsection{Data Analysis}

Quantitative data were analyzed using SPSS (v29). Data preparation included screening for completeness and removal of patterned or low-quality responses. Normality was assessed using Shapiro-Wilk tests. Reliability of the survey scales was evaluated using Cronbach's alpha. Descriptive statistics were calculated for all key variables.

Group comparisons by gender and specialization were conducted using independent-samples t-tests and one-way ANOVA for normally distributed variables, and Mann-Whitney U and Kruskal-Wallis tests for variables violating normality assumptions. This selection ensures that the statistical tests align appropriately with the distributional characteristics of each variable. 

The relationship between risk awareness and adoption willingness was assessed using Pearson or Spearman correlation coefficients, depending on normality. Detailed results of reliability analysis, descriptive statistics, and inferential tests are reported in the Results section.

\section{Results}
\label{sec:results}

\subsection{Demographic Overview}

Table~\ref{tab:demographics} summarizes the demographic characteristics of the final sample ($N = 139$), including age, gender, ethnicity, student status, specialization, years of experience, and self-rated AI knowledge. Descriptive information is further detailed in Section~3.2. These distributions provide important contextual information for interpreting subsequent analyses.

\begin{table}[!ht]
\centering
\caption{Demographic Information of Participants ($N = 139$)}
\label{tab:demographics}
\begin{tabular}{lrr}
\toprule
\textbf{} & \textbf{Count} & \textbf{\%} \\
\midrule
\textbf{Age} & & \\
18--24 years & 88 & 63.3 \\
25--34 years & 48 & 34.5 \\
35--44 years & 3 & 2.2 \\
\midrule
\textbf{Gender} & & \\
Male & 68 & 48.9 \\
Female & 71 & 51.1 \\
\midrule
\textbf{Ethnicity} & & \\
Asian & 108 & 77.7 \\
Black or African American & 3 & 2.2 \\
Hispanic or Latino & 6 & 4.3 \\
White or Caucasian & 15 & 10.8 \\
Mixed or Multiple Ethnicities & 4 & 2.9 \\
Prefer not to say & 3 & 2.2 \\
\midrule
\textbf{Status} & & \\
Current Undergraduate Student & 30 & 21.6 \\
Current Graduate Student & 105 & 75.5 \\
Recent Graduate & 3 & 2.2 \\
Working Professional & 1 & 0.7 \\
\midrule
\textbf{Specialization} & & \\
Computer Science & 48 & 34.5 \\
Data Science / Data Analytics & 68 & 48.9 \\
Other & 23 & 16.5 \\
\midrule
\textbf{Experience} & & \\
Less than 1 year & 39 & 28.1 \\
1--3 years & 66 & 47.5 \\
4--6 years & 30 & 21.6 \\
7+ years & 4 & 2.9 \\
\midrule
\textbf{AI Knowledge} & & \\
Not understanding & 5 & 3.6 \\
Basic understanding & 34 & 24.5 \\
Moderate understanding & 64 & 46.0 \\
Proficient understanding & 27 & 19.4 \\
Expert & 9 & 6.5 \\
\bottomrule
\end{tabular}
\begin{tablenotes}
\small
\item \centering \textit{Note}: Percentages may not sum to 100 due to rounding.
\end{tablenotes}
\end{table}

\subsection{Reliability and Descriptive Statistics}

Table~\ref{tab:reliability_descriptive} reports reliability and descriptive statistics for the three composite scales. All scales demonstrated acceptable to excellent reliability (Explicit: $\alpha = .93$; Scenario-based: $\alpha = .88$; Adoption willingness: $\alpha = .86$). Composite mean scores were used for all subsequent analyses.

Effect sizes (Cohen's $d$ for $t$-tests and $r$ for non-parametric tests) were computed to complement $p$-values and provide estimates of practical significance.

\begin{table}[!ht]
\centering
\small
\caption{Reliability and Descriptive Statistics for Survey Scales}
\label{tab:reliability_descriptive}
\begin{tabular}{lccc}
\toprule
\textbf{Scale} & \textbf{Explicit Risk} & \textbf{Scenario Risk} & \textbf{Adoption Willingness} \\
\midrule
Items & 12 & 10 & 10 \\
Cronbach's $\alpha$ & 0.93 & 0.88 & 0.86 \\
Mean (SD) & 3.70 (0.29) & 3.10 (0.35) & 3.40 (0.28) \\
Item mean range & 2.99--4.01 & 2.39--3.46 & 2.95--3.86 \\
Inter-item correlations & 0.30--0.76 & 0.18--0.66 & 0.09--0.64 \\
Item-total correlations & 0.50--0.77 & 0.48--0.70 & 0.39--0.68 \\
\bottomrule
\end{tabular}
\end{table}

\subsection{Comparative Analysis}

Comparisons were conducted across gender and academic specialization for explicit awareness, scenario-based awareness, and adoption willingness.

\subsubsection{Explicit AI Risk Awareness}
\begin{figure}[!ht]
\centering
\includegraphics[width=0.8\linewidth]{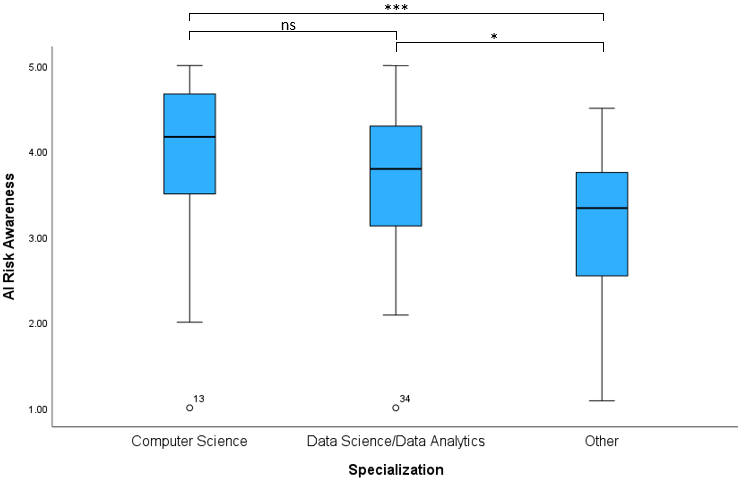}
\caption{Explicit AI Risk Awareness by Academic Specialization. CS and DS/DA students showed significantly higher scores than the Other group.}
\label{fig:boxplot}
\end{figure}

Descriptive patterns across risk categories by gender and specialization are presented in Table~\ref{tab:awareness}. Visual patterns of specialization differences are further illustrated in Figure~\ref{fig:boxplot}, which highlights higher explicit risk awareness among CS and DS/DA students relative to the Other group.

Explicit AI risk awareness showed no gender difference, $U = 2315.50$, $p = .678$, $d = 0.14$, 95\% CI $[-0.198,\, 0.468]$.

Specialization differences were significant, Kruskal--Wallis $\chi^2(2) = 13.91$, $p < .001$. 
Post-hoc comparisons indicated: CS $>$ Other ($p < .001$, $r = .41$, 95\% CI $[.018,\, .191]$); 
DS/DA $>$ Other ($p = .014$, $r = .26$, 95\% CI $[.004,\, .179]$).

\begin{table}[!ht]
\centering
\caption{Mean (SD) of Explicit AI Risk Awareness by Gender and Specialization ($N = 139$)}
\label{tab:awareness}
\small
\resizebox{\textwidth}{!}{
\begin{tabular}{lcccccc}
\toprule
\textbf{Risk Category} & \textbf{Male} & \textbf{Female} & \textbf{CS} & \textbf{DS/DA} & \textbf{Other} & \textbf{Total} \\
\midrule
General AI risks & 3.04 (1.20) & 2.93 (1.19) & 3.02 (1.30) & 3.09 (1.09) & 2.61 (1.23) & 2.99 (1.19) \\
Privacy/data security & 4.19 (1.03) & 3.77 (1.31) & 4.25 (1.18) & 3.97 (1.04) & 3.43 (1.50) & 3.98 (1.20) \\
Bias/fairness & 3.35 (1.09) & 3.39 (1.29) & 3.56 (1.40) & 3.38 (1.05) & 2.96 (1.07) & 3.37 (1.19) \\
Security vulnerabilities & 3.87 (1.11) & 3.80 (1.20) & 4.10 (1.13) & 3.84 (1.09) & 3.26 (1.21) & 3.83 (1.15) \\
Job displacement & 3.76 (1.28) & 3.62 (1.35) & 4.08 (1.15) & 3.62 (1.27) & 3.09 (1.54) & 3.69 (1.31) \\
Misinformation & 3.91 (1.12) & 3.99 (1.17) & 4.08 (1.13) & 3.87 (1.18) & 3.91 (1.04) & 3.95 (1.14) \\
Autonomous weapons & 4.06 (1.05) & 3.58 (1.33) & 4.08 (1.20) & 3.99 (1.07) & 2.74 (1.14) & 3.81 (1.22) \\
Transparency issues & 3.90 (1.05) & 3.72 (1.20) & 4.10 (1.04) & 3.79 (1.11) & 3.22 (1.17) & 3.81 (1.13) \\
Ethical implications & 3.84 (1.03) & 3.73 (1.16) & 4.10 (1.04) & 3.76 (1.02) & 3.17 (1.19) & 3.78 (1.10) \\
Psychological impacts & 3.65 (1.28) & 3.61 (1.35) & 4.04 (1.22) & 3.51 (1.26) & 3.09 (1.41) & 3.63 (1.31) \\
Social inequalities & 3.53 (1.13) & 3.56 (1.30) & 3.96 (1.18) & 3.46 (1.13) & 2.96 (1.26) & 3.55 (1.21) \\
Misuse/abuse & 4.04 (1.23) & 3.99 (1.25) & 4.33 (1.14) & 4.03 (1.17) & 3.30 (1.36) & 4.01 (1.23) \\
\bottomrule
\end{tabular}
}
\end{table}

\subsubsection{Scenario-Based AI Risk Awareness}

Scenario-based risk ratings for each application context are shown in Table~\ref{tab:implicit_awareness}. These results indicate substantial variability across scenarios but limited differences across demographic groups. \par

Gender difference: $t(137) = -0.23$, $p = .819$, $d = -0.06$, 95\% CI $[-0.390,\, 0.275]$. \par
Specialization: non-significant, Kruskal--Wallis $\chi^2(2) = 3.83$, $p = .148$, $\eta^2 = .028$, 95\% CI $[-0.015,\, 0.078]$.

\begin{table}[!ht]
\centering
\caption{Mean (SD) of Scenario-Based AI Risk Awareness by Gender and Specialization ($N = 139$)}
\label{tab:implicit_awareness}
\small
\resizebox{\textwidth}{!}{
\begin{tabular}{lcccccc}
\toprule
\textbf{Risk Category} & Male & Female & CS & DS/DA & Other & Total \\
\midrule
Personal privacy & 2.96 (1.43) & 3.32 (1.20) & 3.13 (1.48) & 3.34 (1.30) & 2.22 (1.09) & 3.14 (1.33) \\
Bias & 3.12 (1.47) & 3.08 (1.42) & 3.25 (1.60) & 3.07 (1.34) & 2.87 (1.39) & 3.10 (1.44) \\
Security vulnerability & 3.09 (1.32) & 3.06 (1.34) & 3.35 (1.36) & 2.81 (1.34) & 3.26 (1.10) & 3.07 (1.33) \\
Job displacement & 2.79 (1.40) & 2.85 (1.25) & 3.02 (1.42) & 2.76 (1.33) & 2.57 (1.04) & 2.82 (1.32) \\
Misinformation & 3.41 (1.40) & 3.18 (1.41) & 3.75 (1.28) & 3.07 (1.40) & 3.00 (1.48) & 3.29 (1.40) \\
Autonomous weapons & 3.40 (1.39) & 3.49 (1.42) & 3.54 (1.44) & 3.32 (1.32) & 3.61 (1.41) & 3.45 (1.37) \\
Transparency & 3.04 (1.40) & 3.28 (1.39) & 3.38 (1.38) & 2.85 (1.39) & 3.65 (1.27) & 3.17 (1.39) \\
Misuse/abuse & 2.76 (1.39) & 3.03 (1.43) & 2.88 (1.44) & 2.63 (1.38) & 3.74 (1.36) & 2.90 (1.41) \\
Psychological impacts & 2.44 (1.30) & 2.32 (1.19) & 2.38 (1.33) & 2.29 (1.23) & 2.65 (1.07) & 2.38 (1.24) \\
Social inequalities & 2.47 (1.38) & 2.39 (1.22) & 2.69 (1.52) & 2.21 (1.17) & 2.57 (1.08) & 2.43 (1.30) \\
\midrule
Composite Score & 2.96 (0.99) & 3.00 (0.91) & 3.14 (1.05) & 2.83 (0.90) & 3.11 (0.83) & 2.98 (0.95) \\
\bottomrule
\end{tabular}}
\end{table}

\subsubsection{Scenario-Based Adoption Willingness}

\begin{figure}[!ht]
\centering
\includegraphics[width=0.95\textwidth]{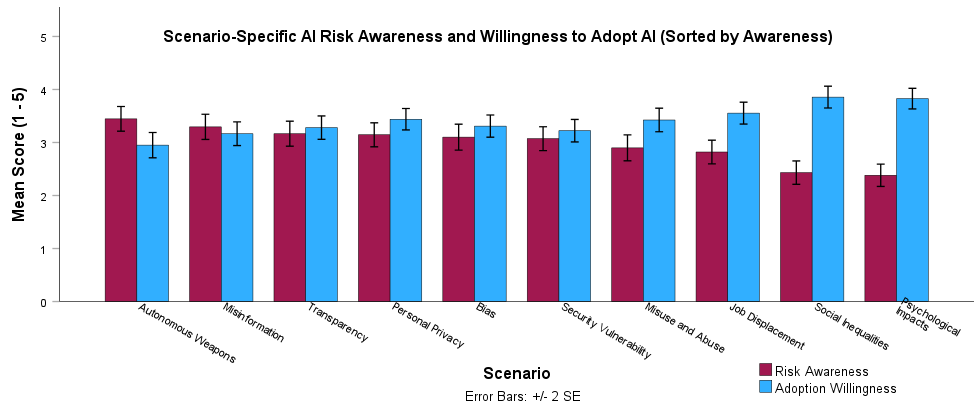}
\caption{Scenario-Based Risk Awareness and Adoption Willingness Across Application Contexts. 
A clear inverse relationship appears: scenarios rated as higher risk consistently show lower willingness to adopt, whereas lower-risk scenarios show higher willingness.}
\label{fig:scenario_comparison_clean}
\end{figure}

Figure~\ref{fig:scenario_comparison_clean} provides a comparison of 
scenario-specific risk awareness and adoption willingness across all ten applications, illustrating the inverse trend between perceived risk and willingness to adopt.

\begin{figure}[!ht]
\centering
\includegraphics[width=0.8\textwidth]{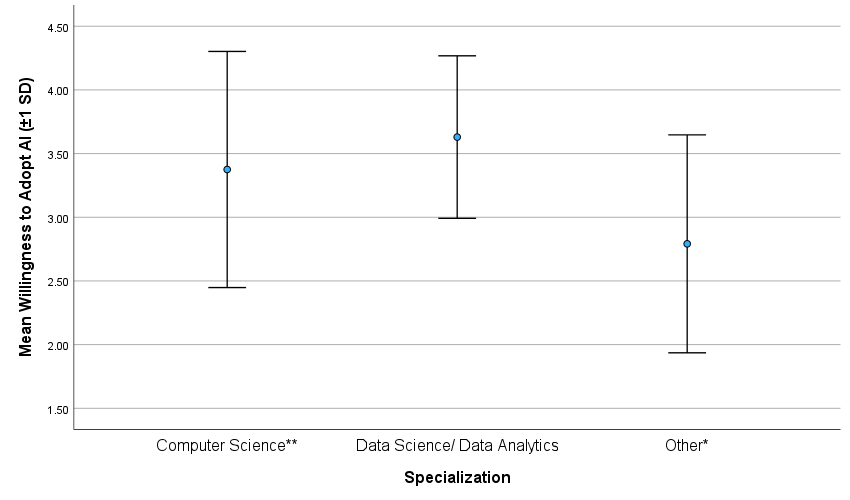}
\caption{Scenario-Based Adoption Willingness by Academic Specialization. CS and DS/DA students reported significantly higher willingness than students in Other disciplines.}
\label{fig:barchart_willingness}
\end{figure}

Scenario-level willingness-to-adopt scores are reported in Table~\ref{tab:scenario_risks}. Figure~\ref{fig:barchart_willingness} provides a visualization of group differences, demonstrating that CS and DS/DA students exhibit notably higher adoption willingness than students from other fields.

Gender difference present: $t(137)=2.02$, $p=.046$, $d=0.34$, 95\% CI $[0.007,\, 0.677]$. \par
Specialization differences significant:
ANOVA $F(2,136)=9.85$, $p<.001$, $\eta^2=.127$, 95\% CI $[.035,\, .226]$.

\begin{table}[!ht]
\centering
\caption{Scenario-Based AI Risk Awareness and Adoption Willingness ($N = 139$)}
\label{tab:scenario_risks}
\begin{tabular}{lcccc}
\toprule
\textbf{Risk Type} & \multicolumn{2}{c}{Risk Awareness} & \multicolumn{2}{c}{Adoption Willingness} \\
 & Mean & SD & Mean & SD \\
\midrule
Personal Privacy & 3.14 & 1.33 & 3.44 & 1.19 \\
Bias & 3.10 & 1.44 & 3.31 & 1.23 \\
Security Vulnerability & 3.07 & 1.33 & 3.22 & 1.25 \\
Job displacement & 2.82 & 1.32 & 3.55 & 1.22 \\
Misinformation & 3.29 & 1.40 & 3.17 & 1.32 \\
Autonomous weapons & 3.45 & 1.37 & 2.95 & 1.41 \\
Transparency & 3.17 & 1.39 & 3.28 & 1.30 \\
Misuse/abuse & 2.90 & 1.44 & 3.42 & 1.31 \\
Psychological impacts & 2.38 & 1.24 & 3.83 & 1.15 \\
Social inequalities & 2.43 & 1.30 & 3.86 & 1.21 \\
\bottomrule
\end{tabular}
\end{table}

\subsubsection{Explicit vs.\ Scenario-Based AI Risk Awareness}

Results of the Wilcoxon signed-rank tests comparing explicit and scenario-based risk ratings are presented in Table~\ref{tab:wilcoxon_results}. These findings confirm consistent and statistically significant differences between explicit and contextualized evaluations across nearly all matched domains.

Explicit awareness ($M = 3.70$) significantly exceeded scenario-based awareness ($M = 2.98$) for nearly all risks, with largest effects for psychological impacts and social inequalities.

\begin{table}[!ht]
\centering
\caption{Wilcoxon Signed-Rank Test: Explicit vs. Scenario-Based AI Risk Awareness}
\label{tab:wilcoxon_results}
\begin{tabular}{lcc}
\toprule
\textbf{Risk/Scenario} & \textbf{Z} & \textbf{$p$} \\
\midrule
Privacy in automation & -5.409 & $<.001$ \\
Bias in hiring & -1.822 & .068 \\
Security in banking & -5.074 & $<.001$ \\
Job displacement in manufacturing & -5.858 & $<.001$ \\
Manipulation in social media & -4.616 & $<.001$ \\
Autonomous weapons in drones & -2.211 & .027 \\
Transparency in healthcare & -3.980 & $<.001$ \\
Misuse in mass surveillance & -6.061 & $<.001$ \\
Psychological impacts of assistants & -7.200 & $<.001$ \\
Social inequalities in education & -6.730 & $<.001$ \\
\bottomrule
\end{tabular}
\end{table}

\section{Discussion}

Before interpreting the findings, it is important to note that the study's cross-sectional design permits only correlational inferences; causal relationships between AI risk perception and adoption willingness cannot be established.

\subsection{Key Findings}

\subsubsection{Explicit Risks Elicit Stronger Concern than Scenario-Based Risks}\label{kf1}
The results reveal a critical divergence in technology students' AI risk perception: explicitly stated risks elicit substantially stronger concern than risks embedded within specific application scenarios. A clear example appears in the domain of privacy: students rated explicit privacy/data security risks highly ($M = 3.98$), yet their concern declined markedly ($M = 3.14$) when the same risk was embedded in a realistic smart-home scenario. Across the full sample, the overall mean of explicit AI risk ratings was $3.70$, notably higher than the $3.10$ observed for scenario-based AI risk ratings.

These findings suggest that students recognize and articulate AI risks more strongly when risks are directly named and conceptually framed, a pattern consistent with research showing that abstract descriptions can activate higher-level evaluative reasoning \citep{trope2010construal}. However, this awareness weakens when those same risks are presented in concrete, operational contexts. In short, explicit risk knowledge does not fully transfer to contextualized risk recognition, highlighting a gap between abstract understanding and applied judgment.

\subsubsection{Perceived Risk and Adoption Willingness Are Inversely Related}\label{kf2}
The results provide strong evidence for an inverse relationship between perceived contextual risk and willingness to adopt AI technologies. Across all scenarios, students showed the lowest adoption willingness in domains they perceived as high-risk---most notably autonomous weapons ($M_{\text{risk}} = 3.45$, $M_{\text{adopt}} = 2.95$) and misinformation/manipulation ($M_{\text{risk}} = 3.29$, $M_{\text{adopt}} = 3.17$). In contrast, the highest adoption willingness occurred in domains perceived as lower-risk, such as psychological impacts ($M_{\text{risk}} = 2.38$, $M_{\text{adopt}} = 3.83$) and social inequalities ($M_{\text{risk}} = 2.43$, $M_{\text{adopt}} = 3.86$). This consistent pattern demonstrates that adoption judgments are shaped less by general attitudes toward AI and more by scenario-specific risk cues.

Taken together, the findings show that lower perceived risk reliably predicts greater willingness to adopt AI, even within the same student population and using parallelized scenario formats. This inverse association highlights the need for AI literacy and ethics education to incorporate contextualized risk assessments rather than relying solely on general or abstract discussions of AI safety.

\subsubsection{Education Equalizes Risk Awareness but Gender Differences Persist in Adoption}\label{kf3}

The results show that formal technical education effectively equalizes gender differences in AI risk perception. Male and female students demonstrated statistically equivalent levels of explicit risk awareness ($p = .678$) and scenario-based risk awareness ($p = .819$), indicating that domain-specific instruction provides a shared framework for evaluating AI-related harms.

Despite this convergence in risk perception, a clear behavioral divergence emerged. Male students reported significantly higher willingness to adopt AI technologies than female students ($M_{\text{male}} = 3.55$ vs.\ $M_{\text{female}} = 3.26$, $p = .046$, $d = 0.34$). This suggests that equalizing cognitive understanding of risks does not necessarily translate into equal adoption behavior.

This dissociation between awareness and behavior aligns with prior research showing that gender differences in uncertainty tolerance, trust, and perceived behavioral control can influence technology-related decisions \citep{venkatesh2003user,byrnes1999gender,gefen1997gender}. Taken together, these findings highlight that while technical education may narrow gender gaps in evaluative judgments, it does not eliminate underlying psychological or sociocultural factors that shape willingness to adopt AI. As such, AI literacy initiatives should complement technical instruction with strategies that address these behavioral dimensions to support equitable engagement with AI technologies.

\subsubsection{Specialization Differences Reveal Evidence of Applied Risk Underappreciation}\label{kf4}

A clear pattern emerged when comparing AI-related specializations with students from non-technical fields. Computer Science (CS) and Data Science/Data Analytics (DS/DA) students demonstrated higher explicit AI risk awareness than students in the ``Other'' group, consistent with their stronger curricular exposure to AI systems. However, these same CS and DS/DA students did not show correspondingly higher recognition of risks when evaluating scenario-based applications, where their average scenario-awareness scores were comparable to or only marginally higher than those of non-technical peers.

Despite this parallel in scenario-based risk perception, CS and DS/DA students expressed substantially greater willingness to adopt AI technologies across nearly all application contexts. Students from non-technical fields, by contrast, showed markedly lower adoption willingness, even when their scenario-based risk assessments were similar. This divergence suggests that technical specialization strengthens confidence and perceived control over AI systems, which in turn elevates willingness to adopt AI regardless of contextualized risk cues.

Taken together, these results point to a form of \emph{risk underappreciation} among students in AI-related fields: they exhibit high explicit awareness of AI risks yet demonstrate muted sensitivity to those same risks when embedded in applied scenarios, while simultaneously showing strong enthusiasm for adoption. A plausible mechanism underlying this pattern is that greater domain familiarity can increase perceived control and reduce sensitivity to potential harms, a tendency documented in research showing that expertise often promotes overconfidence and higher risk-taking \citep{hilary2006does,said2023artificial}. The present findings extend this work by providing quantitative evidence of risk underappreciation within the context of university-level AI education.

\subsection{Implications and Practical Recommendations}

The four key findings of this study collectively reveal several important implications for AI education, curriculum design, and responsible AI practice. These implications extend beyond the empirical patterns reported earlier and point to concrete opportunities for strengthening students' ability to evaluate and engage with AI systems responsibly.

In addition to these empirical findings, the study also contributes to broader work on AI risk perception and AI literacy. First, it offers a combined analysis of explicit and scenario-based AI risk awareness within the same population, enabling a direct comparison of articulated versus contextualized judgments—an approach that remains relatively uncommon in existing research. Second, it provides empirical evidence of how perceived contextual risk relates to adoption willingness across different application domains. Third, it highlights how gender and disciplinary identity shape adoption behavior even when risk awareness itself is comparable. Together, these contributions help refine theoretical understanding of AI risk perception and guide the development of more effective AI risk education strategies.

\subsubsection*{1. Strengthening Transfer Between Abstract Risk Knowledge and Applied Contexts}

Section~\ref{kf1} showed that students recognize AI risks more strongly when they are explicitly labeled than when the same risks are embedded in realistic scenarios. This gap suggests that existing AI curricula may succeed in conveying conceptual terminology but fall short in preparing students to detect risks in operational settings. To address this disconnect, educational programs should integrate scenario-based reasoning throughout instruction---for example, by embedding case analyses, incident walkthroughs, and structured reflection exercises that require students to identify risks without being prompted by explicit labels. These strategies can support the transfer of abstract risk knowledge into applied judgment, a capacity essential for responsible AI development.

\subsubsection*{2. Embedding Contextualized Risk Evaluation Into AI Literacy}

Section~\ref{kf2} demonstrated that adoption willingness closely follows contextual risk perception, highlighting the importance of teaching students how to evaluate risks not just at the conceptual level but within specific application domains. To cultivate more calibrated decision-making, AI ethics modules should incorporate domain-specific exemplars across health, security, education, entertainment, and workplace contexts. Asking students to articulate both the potential harms and justifications for adoption in each context can strengthen their ability to weigh domain-relevant trade-offs and avoid over-reliance on general attitudes toward AI.

\subsubsection*{3. Addressing Behavioral Disparities Through Psychological and Sociocultural Support}

Although section~\ref{kf3} showed that technical education equalized gender differences in risk perception, it also revealed that gender gaps persist in adoption behaviors. This indicates that cognitive understanding alone is insufficient for equitable engagement. AI curricula should therefore integrate interventions that address psychological and sociocultural factors shaping adoption decisions. Activities that build self-efficacy, reduce uncertainty aversion, and strengthen calibrated trust in AI systems may help reduce gender-linked behavioral disparities. Importantly, such interventions should be framed not as remediation but as part of a broader commitment to supporting diverse forms of engagement with AI.

\subsubsection*{4. Expanding AI Literacy Beyond Technical Majors to Counteract Overconfidence and Vulnerability}

Section~\ref{kf4} highlighted a dual challenge: students in AI-related majors demonstrate strong explicit awareness yet show muted sensitivity to contextualized risks and elevated willingness to adopt AI technologies. Meanwhile, students in non-technical fields show lower adoption willingness despite comparable scenario-based awareness. These patterns indicate that both groups would benefit from tailored educational strategies. For technical majors, instruction should emphasize critical self-reflection and highlight the limits of expertise to reduce overconfidence and mitigate forms of risk underappreciation. For non-technical students, AI literacy programs should focus on building foundational understanding and confidence, ensuring they are not marginalized in technology discourse or decision-making.

\subsubsection*{5. Building Institution-Level Frameworks for Responsible AI Education}

Across section~\ref{kf1}--~\ref{kf4}, a consistent theme emerges: students require structured opportunities to integrate conceptual understanding, contextual reasoning, and reflective judgment. Institutions can support this integration by adopting program-wide frameworks that map AI risk competencies across curricula, ensuring that explicit instruction, scenario-based reasoning, and reflective activities appear throughout a student's academic trajectory. Periodic needs assessments, informed by student feedback and evolving AI technologies, can help institutions adapt AI risk education to diverse learner profiles and emerging societal challenges.

\subsection{Limitations of the Study}

While this study provides meaningful insights into technology students' AI risk awareness and adoption behavior, several limitations should be acknowledged to contextualize the findings.

First, the data were collected from a single private university in the United States, which limits generalizability beyond similar institutional and disciplinary contexts. Participation was voluntary, which may have attracted students with higher interest in or familiarity with AI. As a result, the levels of risk awareness and adoption willingness observed in this sample may not fully reflect those of the broader student population.

Second, although the sample included students from Computer Science, Data Science/Data Analytics, and a range of other majors, the ``Other'' group was comparatively small. This limits the ability to draw strong conclusions about non-technical fields and constrains the interpretation of disciplinary differences. Future research with more balanced representation across academic disciplines would strengthen the robustness of these comparisons.

Third, the study relied on self-reported measures of risk perception, self-assessed AI knowledge, and adoption intentions. Such measures are susceptible to social desirability bias and may not fully correspond to actual behavior or objective proficiency. Incorporating performance-based assessments, behavioral tasks, or multimodal measures would help validate and extend the present findings.

Fourth, the cross-sectional design restricts interpretation to correlational relationships. Because all variables were measured at a single time point, it is not possible to infer causal pathways between AI risk perception and adoption willingness, nor to examine how these perceptions evolve as students progress through their academic programs or gain professional experience. Longitudinal or repeated-measures studies would allow for analysis of developmental trajectories and provide stronger evidence on the impact of educational interventions.

These limitations highlight the need for cautious interpretation of the findings and indicate clear opportunities for future research to build on and expand the present work.

\section{Conclusion}

As AI technologies become increasingly integrated into society, understanding how future professionals evaluate AI-related risks is essential. This study examined technology students' explicit and scenario-based AI risk awareness, their adoption willingness across application domains, and the influence of gender and disciplinary background on these judgments.

Four consistent patterns emerged. First, students expressed substantially stronger concern for explicitly stated AI risks than for the same risks embedded in realistic scenarios, indicating that abstract knowledge does not readily translate into contextual risk recognition. Second, adoption willingness declined systematically as perceived contextual risk increased, demonstrating that students' acceptance of AI systems is closely tied to their appraisal of domain-specific risks. Third, although technical training appeared to equalize gender differences in risk awareness, male students consistently reported higher willingness to adopt AI technologies, revealing a behavioral gap not explained by differences in evaluative judgment. Fourth, students in AI-intensive majors such as Computer Science and Data Science showed both higher explicit risk awareness and higher adoption willingness than their non-technical peers—a pattern consistent with the form of ``risk underappreciation'' in which technical familiarity coexists with reduced caution in applied contexts.

Together, these findings show that technology students' engagement with AI is shaped by both how risks are framed and who is evaluating them. Strengthening responsible AI practice will therefore require educational approaches that pair explicit instruction with scenario-based reasoning, address behavioral and psychological dimensions of adoption, and support more calibrated decision-making across disciplinary groups. Such efforts will be critical as AI continues to expand across academic and professional environments.





\section*{Funding Statement}
This research did not receive any specific grant from funding agencies in the public, commercial, or not-for-profit sectors.

\section*{Declaration of AI Assisted Writing} During the preparation of this work the authors used ChatGPT (OpenAI) in order to improve language and readability. After using this tool, the authors reviewed and edited the content as needed and take full responsibility for the content of the publication.

\section*{Data Availability} 
The datasets generated and analyzed during the current study are not publicly available due to participant confidentiality but are available from the corresponding author on reasonable request.

\section*{Ethics Approval and Consent to Participate}

This study was approved by the Institutional Review Board at Clark University (Protocol No. 701). All participants provided informed consent prior to participation.

\bibliographystyle{elsarticle-harv}  
\bibliography{reference}

\begin{thebibliography}{64}
\expandafter\ifx\csname natexlab\endcsname\relax\def\natexlab#1{#1}\fi
\providecommand{\url}[1]{\texttt{#1}}
\providecommand{\href}[2]{#2}
\providecommand{\path}[1]{#1}
\providecommand{\DOIprefix}{doi:}
\providecommand{\ArXivprefix}{arXiv:}
\providecommand{\URLprefix}{URL: }
\providecommand{\Pubmedprefix}{pmid:}
\providecommand{\doi}[1]{\href{http://dx.doi.org/#1}{\path{#1}}}
\providecommand{\Pubmed}[1]{\href{pmid:#1}{\path{#1}}}
\providecommand{\bibinfo}[2]{#2}
\ifx\xfnm\relax \def\xfnm[#1]{\unskip,\space#1}\fi
\bibitem[{Agarwal et~al.(2020)Agarwal, Farid, El-Gaaly and
  Lim}]{agarwal2020detecting}
\bibinfo{author}{Agarwal, S.}, \bibinfo{author}{Farid, H.},
  \bibinfo{author}{El-Gaaly, T.}, \bibinfo{author}{Lim, S.N.},
  \bibinfo{year}{2020}.
\newblock \bibinfo{title}{Detecting deep-fake videos from appearance and
  behavior}, in: \bibinfo{booktitle}{2020 IEEE International Workshop on
  Information Forensics and Security (WIFS)}, pp. \bibinfo{pages}{1--6}.
\newblock \DOIprefix\doi{10.1109/WIFS49906.2020.9360904}.
\bibitem[{A{\"\i}meur et~al.(2023)A{\"\i}meur, Amri and
  Brassard}]{aimeur2023fake}
\bibinfo{author}{A{\"\i}meur, E.}, \bibinfo{author}{Amri, S.},
  \bibinfo{author}{Brassard, G.}, \bibinfo{year}{2023}.
\newblock \bibinfo{title}{Fake news, disinformation and misinformation in
  social media: a review}.
\newblock \bibinfo{journal}{Social Network Analysis and Mining}
  \bibinfo{volume}{13}, \bibinfo{pages}{30}.
\newblock \URLprefix \url{https://doi.org/10.1007/s13278-023-01028-5},
  \DOIprefix\doi{10.1007/s13278-023-01028-5}.
\bibitem[{Ajzen(1991)}]{ajzen1991theory}
\bibinfo{author}{Ajzen, I.}, \bibinfo{year}{1991}.
\newblock \bibinfo{title}{The theory of planned behavior}.
\newblock \bibinfo{journal}{Organizational Behavior and Human Decision
  Processes} \bibinfo{volume}{50}, \bibinfo{pages}{179--211}.
\newblock \URLprefix
  \url{https://www.sciencedirect.com/science/article/pii/074959789190020T},
  \DOIprefix\doi{10.1016/0749-5978(91)90020-T}. \bibinfo{note}{theories of
  Cognitive Self-Regulation}.
\bibitem[{Al-Asadi and Tasdemir(2021)}]{al2021using}
\bibinfo{author}{Al-Asadi, M.A.}, \bibinfo{author}{Tasdemir, S.},
  \bibinfo{year}{2021}.
\newblock \bibinfo{title}{Using artificial intelligence against the phenomenon
  of fake news: A systematic literature review}.
\newblock \bibinfo{journal}{Combating Fake News with Computational Intelligence
  Techniques} , \bibinfo{pages}{39--54}\URLprefix
  \url{https://doi.org/10.1007/978-3-030-90087-8_2},
  \DOIprefix\doi{10.1007/978-3-030-90087-8_2}.
\bibitem[{Albaroudi et~al.(2024)Albaroudi, Mansouri and
  Alameer}]{albaroudi2024comprehensive}
\bibinfo{author}{Albaroudi, E.}, \bibinfo{author}{Mansouri, T.},
  \bibinfo{author}{Alameer, A.}, \bibinfo{year}{2024}.
\newblock \bibinfo{title}{A comprehensive review of ai techniques for
  addressing algorithmic bias in job hiring}.
\newblock \bibinfo{journal}{AI} \bibinfo{volume}{5}, \bibinfo{pages}{383--404}.
\newblock \URLprefix \url{https://www.mdpi.com/2673-2688/5/1/19},
  \DOIprefix\doi{10.3390/ai5010019}.
\bibitem[{Anderljung et~al.(2024)Anderljung, Hazell and
  Von~Knebel}]{anderljung2024protecting}
\bibinfo{author}{Anderljung, M.}, \bibinfo{author}{Hazell, J.},
  \bibinfo{author}{Von~Knebel, M.}, \bibinfo{year}{2024}.
\newblock \bibinfo{title}{Protecting society from {AI} misuse: when are
  restrictions on capabilities warranted?}
\newblock \bibinfo{journal}{AI \& Society} , \bibinfo{pages}{1--17}\URLprefix
  \url{https://doi.org/10.1007/s00146-024-02130-8},
  \DOIprefix\doi{10.1007/s00146-024-02130-8}.
\bibitem[{Archambault et~al.(2024)Archambault, Ramachandran, Acosta and
  Fu}]{archambault2024ethical}
\bibinfo{author}{Archambault, S.G.}, \bibinfo{author}{Ramachandran, S.},
  \bibinfo{author}{Acosta, E.}, \bibinfo{author}{Fu, S.}, \bibinfo{year}{2024}.
\newblock \bibinfo{title}{Ethical dimensions of algorithmic literacy for
  college students: Case studies and cross-disciplinary connections}.
\newblock \bibinfo{journal}{The Journal of Academic Librarianship}
  \bibinfo{volume}{50}, \bibinfo{pages}{102865}.
\newblock \URLprefix
  \url{https://www.sciencedirect.com/science/article/pii/S0099133324000260},
  \DOIprefix\doi{10.1016/j.acalib.2024.102865}.
\bibitem[{Benzinger et~al.(2023)Benzinger, Ursin, Balke, Kacprowski and
  Salloch}]{benzinger2023should}
\bibinfo{author}{Benzinger, L.}, \bibinfo{author}{Ursin, F.},
  \bibinfo{author}{Balke, W.T.}, \bibinfo{author}{Kacprowski, T.},
  \bibinfo{author}{Salloch, S.}, \bibinfo{year}{2023}.
\newblock \bibinfo{title}{Should artificial intelligence be used to support
  clinical ethical decision-making? {A} systematic review of reasons}.
\newblock \bibinfo{journal}{BMC Medical Ethics} \bibinfo{volume}{24},
  \bibinfo{pages}{48}.
\newblock \URLprefix \url{https://doi.org/10.1186/s12910-023-00929-6},
  \DOIprefix\doi{10.1186/s12910-023-00929-6}.
\bibitem[{Bernal and Mazo(2022)}]{bernal2022transparency}
\bibinfo{author}{Bernal, J.}, \bibinfo{author}{Mazo, C.}, \bibinfo{year}{2022}.
\newblock \bibinfo{title}{Transparency of artificial intelligence in
  healthcare: Insights from professionals in computing and healthcare
  worldwide}.
\newblock \bibinfo{journal}{Applied Sciences} \bibinfo{volume}{12}.
\newblock \URLprefix \url{https://www.mdpi.com/2076-3417/12/20/10228},
  \DOIprefix\doi{10.3390/app122010228}.
\bibitem[{Brauner et~al.(2023)Brauner, Hick, Philipsen and
  Ziefle}]{brauner2023does}
\bibinfo{author}{Brauner, P.}, \bibinfo{author}{Hick, A.},
  \bibinfo{author}{Philipsen, R.}, \bibinfo{author}{Ziefle, M.},
  \bibinfo{year}{2023}.
\newblock \bibinfo{title}{What does the public think about artificial
  intelligence?—a criticality map to understand bias in the public perception
  of ai}.
\newblock \bibinfo{journal}{Frontiers in Computer Science}
  \bibinfo{volume}{Volume 5 - 2023}.
\newblock \URLprefix
  \url{https://www.frontiersin.org/journals/computer-science/articles/10.3389/fcomp.2023.1113903},
  \DOIprefix\doi{10.3389/fcomp.2023.1113903}.
\bibitem[{Brown et~al.(2024)Brown, Xie, Sarder, Fiesler and
  Wiese}]{brown2024teaching}
\bibinfo{author}{Brown, N.}, \bibinfo{author}{Xie, B.},
  \bibinfo{author}{Sarder, E.}, \bibinfo{author}{Fiesler, C.},
  \bibinfo{author}{Wiese, E.S.}, \bibinfo{year}{2024}.
\newblock \bibinfo{title}{Teaching ethics in computing: A systematic literature
  review of acm computer science education publications}.
\newblock \bibinfo{journal}{ACM Trans. Comput. Educ.} \bibinfo{volume}{24}.
\newblock \URLprefix \url{https://doi.org/10.1145/3634685},
  \DOIprefix\doi{10.1145/3634685}.
\bibitem[{Byrnes et~al.(1999)Byrnes, Miller and Schafer}]{byrnes1999gender}
\bibinfo{author}{Byrnes, J.P.}, \bibinfo{author}{Miller, D.C.},
  \bibinfo{author}{Schafer, W.D.}, \bibinfo{year}{1999}.
\newblock \bibinfo{title}{Gender differences in risk taking: A meta-analysis}.
\newblock \bibinfo{journal}{Psychological Bulletin} \bibinfo{volume}{125},
  \bibinfo{pages}{367--383}.
\newblock \URLprefix
  \url{https://psycnet.apa.org/fulltext/1999-13573-004.html}.
\bibitem[{Capraro et~al.(2024)Capraro, Lentsch, Acemoglu, Akgun, Akhmedova,
  Bilancini, Bonnefon, Brañas-Garza, Butera, Douglas, Everett, Gigerenzer,
  Greenhow, Hashimoto, Holt-Lunstad, Jetten, Johnson, Kunz, Longoni, Lunn,
  Natale, Paluch, Rahwan, Selwyn, Singh, Suri, Sutcliffe, Tomlinson, van~der
  Linden, Van~Lange, Wall, Van~Bavel and Viale}]{capraro2024impact}
\bibinfo{author}{Capraro, V.}, \bibinfo{author}{Lentsch, A.},
  \bibinfo{author}{Acemoglu, D.}, \bibinfo{author}{Akgun, S.},
  \bibinfo{author}{Akhmedova, A.}, \bibinfo{author}{Bilancini, E.},
  \bibinfo{author}{Bonnefon, J.F.}, \bibinfo{author}{Brañas-Garza, P.},
  \bibinfo{author}{Butera, L.}, \bibinfo{author}{Douglas, K.M.},
  \bibinfo{author}{Everett, J.A.C.}, \bibinfo{author}{Gigerenzer, G.},
  \bibinfo{author}{Greenhow, C.}, \bibinfo{author}{Hashimoto, D.A.},
  \bibinfo{author}{Holt-Lunstad, J.}, \bibinfo{author}{Jetten, J.},
  \bibinfo{author}{Johnson, S.}, \bibinfo{author}{Kunz, W.H.},
  \bibinfo{author}{Longoni, C.}, \bibinfo{author}{Lunn, P.},
  \bibinfo{author}{Natale, S.}, \bibinfo{author}{Paluch, S.},
  \bibinfo{author}{Rahwan, I.}, \bibinfo{author}{Selwyn, N.},
  \bibinfo{author}{Singh, V.}, \bibinfo{author}{Suri, S.},
  \bibinfo{author}{Sutcliffe, J.}, \bibinfo{author}{Tomlinson, J.},
  \bibinfo{author}{van~der Linden, S.}, \bibinfo{author}{Van~Lange, P.A.M.},
  \bibinfo{author}{Wall, F.}, \bibinfo{author}{Van~Bavel, J.J.},
  \bibinfo{author}{Viale, R.}, \bibinfo{year}{2024}.
\newblock \bibinfo{title}{The impact of generative artificial intelligence on
  socioeconomic inequalities and policy making}.
\newblock \bibinfo{journal}{PNAS Nexus} \bibinfo{volume}{3},
  \bibinfo{pages}{pgae191}.
\newblock \DOIprefix\doi{10.1093/pnasnexus/pgae191}.
\bibitem[{Chen et~al.(2019)Chen, Liu, Xiang, Niu, Tong and
  Han}]{chen2019adversarial}
\bibinfo{author}{Chen, T.}, \bibinfo{author}{Liu, J.}, \bibinfo{author}{Xiang,
  Y.}, \bibinfo{author}{Niu, W.}, \bibinfo{author}{Tong, E.},
  \bibinfo{author}{Han, Z.}, \bibinfo{year}{2019}.
\newblock \bibinfo{title}{Adversarial attack and defense in reinforcement
  learning-from {AI} security view}.
\newblock \bibinfo{journal}{Cybersecurity} \bibinfo{volume}{2},
  \bibinfo{pages}{1--22}.
\newblock \URLprefix \url{https://doi.org/10.1186/s42400-019-0027-x},
  \DOIprefix\doi{10.1186/s42400-019-0027-x}.
\bibitem[{Crockett et~al.(2020)Crockett, Garratt, Latham, Colyer and
  Goltz}]{crockett2020risk}
\bibinfo{author}{Crockett, K.}, \bibinfo{author}{Garratt, M.},
  \bibinfo{author}{Latham, A.}, \bibinfo{author}{Colyer, E.},
  \bibinfo{author}{Goltz, S.}, \bibinfo{year}{2020}.
\newblock \bibinfo{title}{Risk and trust perceptions of the public of artifical
  intelligence applications}, in: \bibinfo{booktitle}{2020 International Joint
  Conference on Neural Networks (IJCNN)}, pp. \bibinfo{pages}{1--8}.
\newblock \DOIprefix\doi{10.1109/IJCNN48605.2020.9207654}.
\bibitem[{Delello et~al.(2023)Delello, Sung, Mokhtari and
  De~Giuseppe}]{delello2023exploring}
\bibinfo{author}{Delello, J.A.}, \bibinfo{author}{Sung, W.},
  \bibinfo{author}{Mokhtari, K.}, \bibinfo{author}{De~Giuseppe, T.},
  \bibinfo{year}{2023}.
\newblock \bibinfo{title}{Exploring college students' awareness of {AI and
  ChatGPT}: Unveiling perceived benefits and risks}.
\newblock \bibinfo{journal}{Journal of Inclusive Methodology and Technology in
  Learning and Teaching} \bibinfo{volume}{3}, \bibinfo{pages}{1--25}.
\newblock \URLprefix
  \url{https://www.inclusiveteaching.it/index.php/inclusiveteaching/article/view/132},
  \DOIprefix\doi{10.32043/jimtlt.v3i4.132}.
\bibitem[{{European Union}(2024)}]{EUAIAct2024}
\bibinfo{author}{{European Union}}, \bibinfo{year}{2024}.
\newblock \bibinfo{title}{Artificial intelligence act}.
\newblock
  \bibinfo{howpublished}{\url{https://artificialintelligenceact.eu/ai-act-explorer/}}.
\bibitem[{Floridi(2010)}]{floridi2010cambridge}
\bibinfo{author}{Floridi, L.}, \bibinfo{year}{2010}.
\newblock \bibinfo{title}{Ethics after the information revolution}, in:
  \bibinfo{editor}{Floridi, L.} (Ed.), \bibinfo{booktitle}{The Cambridge
  Handbook of Information and Computer Ethics}. \bibinfo{publisher}{Cambridge
  University Press}, \bibinfo{address}{Cambridge}, pp. \bibinfo{pages}{3--19}.
\newblock \URLprefix \url{https://doi.org/10.1017/CBO9780511845239.002},
  \DOIprefix\doi{10.1017/CBO9780511845239.002}.
\bibitem[{Gefen and Straub(1997)}]{gefen1997gender}
\bibinfo{author}{Gefen, D.}, \bibinfo{author}{Straub, D.W.},
  \bibinfo{year}{1997}.
\newblock \bibinfo{title}{Gender differences in the perception and use of
  e-mail: An extension to the technology acceptance model}.
\newblock \bibinfo{journal}{MIS Quarterly} , \bibinfo{pages}{389--400}.
\bibitem[{Ghotbi and Ho(2021)}]{ghotbi2021moral}
\bibinfo{author}{Ghotbi, N.}, \bibinfo{author}{Ho, M.T.}, \bibinfo{year}{2021}.
\newblock \bibinfo{title}{Moral awareness of college students regarding
  artificial intelligence}.
\newblock \bibinfo{journal}{Asian Bioethics Review} \bibinfo{volume}{13},
  \bibinfo{pages}{421--433}.
\newblock \URLprefix \url{https://doi.org/10.1007/s41649-021-00182-2},
  \DOIprefix\doi{10.1007/s41649-021-00182-2}.
\bibitem[{Ángel {Gómez de Ágreda}(2020)}]{de2020ethics}
\bibinfo{author}{Ángel {Gómez de Ágreda}}, \bibinfo{year}{2020}.
\newblock \bibinfo{title}{Ethics of autonomous weapons systems and its
  applicability to any ai systems}.
\newblock \bibinfo{journal}{Telecommunications Policy} \bibinfo{volume}{44},
  \bibinfo{pages}{101953}.
\newblock \URLprefix
  \url{https://www.sciencedirect.com/science/article/pii/S0308596120300458},
  \DOIprefix\doi{10.1016/j.telpol.2020.101953}. \bibinfo{note}{artificial
  intelligence, economy and society}.
\bibitem[{Hilary and Menzly(2006)}]{hilary2006does}
\bibinfo{author}{Hilary, G.}, \bibinfo{author}{Menzly, L.},
  \bibinfo{year}{2006}.
\newblock \bibinfo{title}{Does past success lead analysts to become
  overconfident?}
\newblock \bibinfo{journal}{Review of Financial Studies} \bibinfo{volume}{19},
  \bibinfo{pages}{1--31}.
\newblock \URLprefix \url{https://doi.org/10.1287/mnsc.1050.0485},
  \DOIprefix\doi{10.1287/mnsc.1050.0485}.
\bibitem[{Jeffrey(2020)}]{jeffrey2020understanding}
\bibinfo{author}{Jeffrey, T.}, \bibinfo{year}{2020}.
\newblock \bibinfo{title}{Understanding college student perceptions of
  artificial intelligence}.
\newblock \bibinfo{journal}{Systemics, Cybernetics and Informatics}
  \bibinfo{volume}{18}, \bibinfo{pages}{8--13}.
\newblock \URLprefix
  \url{https://www.iiisci.org/Journal/PDV/sci/pdfs/HB785NN20.pdf}.
\bibitem[{Kahneman and Tversky(2013)}]{kahneman2013prospect}
\bibinfo{author}{Kahneman, D.}, \bibinfo{author}{Tversky, A.},
  \bibinfo{year}{2013}.
\newblock \bibinfo{title}{Prospect theory: An analysis of decision under risk},
  in: \bibinfo{booktitle}{Handbook of the fundamentals of financial decision
  making: Part I}. \bibinfo{publisher}{World Scientific}, pp.
  \bibinfo{pages}{99--127}.
\newblock \URLprefix
  \url{https://www.worldscientific.com/doi/abs/10.1142/9789814417358_0006},
  \DOIprefix\doi{10.1142/9789814417358_0006}.
\bibitem[{Kasinidou et~al.(2021a)Kasinidou, Kleanthous, Barlas and
  Otterbacher}]{kasinidou2021agree}
\bibinfo{author}{Kasinidou, M.}, \bibinfo{author}{Kleanthous, S.},
  \bibinfo{author}{Barlas, P.}, \bibinfo{author}{Otterbacher, J.},
  \bibinfo{year}{2021}a.
\newblock \bibinfo{title}{I agree with the decision, but they didn't deserve
  this: Future developers' perception of fairness in algorithmic decisions},
  in: \bibinfo{booktitle}{Proceedings of the 2021 acm conference on fairness,
  accountability, and transparency}, pp. \bibinfo{pages}{690--700}.
\newblock \URLprefix \url{https://doi.org/10.1145/3442188.3445931},
  \DOIprefix\doi{10.1145/3442188.3445931}.
\bibitem[{Kasinidou et~al.(2021b)Kasinidou, Kleanthous, Orphanou and
  Otterbacher}]{kasinidou2021educating}
\bibinfo{author}{Kasinidou, M.}, \bibinfo{author}{Kleanthous, S.},
  \bibinfo{author}{Orphanou, K.}, \bibinfo{author}{Otterbacher, J.},
  \bibinfo{year}{2021}b.
\newblock \bibinfo{title}{Educating computer science students about algorithmic
  fairness, accountability, transparency and ethics}, in:
  \bibinfo{booktitle}{Proceedings of the 26th ACM Conference on Innovation and
  Technology in Computer Science Education V. 1}, pp.
  \bibinfo{pages}{484--490}.
\newblock \URLprefix \url{https://doi.org/10.1145/3430665.3456311},
  \DOIprefix\doi{10.1145/3430665.3456311}.
\bibitem[{Larsson and Heintz(2020)}]{larsson2020transparency}
\bibinfo{author}{Larsson, S.}, \bibinfo{author}{Heintz, F.},
  \bibinfo{year}{2020}.
\newblock \bibinfo{title}{Transparency in artificial intelligence}.
\newblock \bibinfo{journal}{Internet Policy Review} \bibinfo{volume}{9},
  \bibinfo{pages}{1--16}.
\newblock \URLprefix
  \url{https://lup.lub.lu.se/search/files/79208055/Larsson_Heintz_2020_Transparency_in_artificial_intelligence_2020_05_05.pdf},
  \DOIprefix\doi{10.14763/2020.2.1469}.
\bibitem[{Lee et~al.(2023)Lee, Bubeck and Petro}]{lee2023benefits}
\bibinfo{author}{Lee, P.}, \bibinfo{author}{Bubeck, S.},
  \bibinfo{author}{Petro, J.}, \bibinfo{year}{2023}.
\newblock \bibinfo{title}{Benefits, limits, and risks of {GPT-4 as an AI}
  chatbot for medicine}.
\newblock \bibinfo{journal}{New England Journal of Medicine}
  \bibinfo{volume}{388}, \bibinfo{pages}{1233--1239}.
\newblock \URLprefix \url{https://www.nejm.org/doi/full/10.1056/NEJMsr2214184},
  \DOIprefix\doi{10.1056/NEJMsr2214184},
  \href{http://arxiv.org/abs/https://www.nejm.org/doi/pdf/10.1056/NEJMsr2214184}{{\tt
  arXiv:https://www.nejm.org/doi/pdf/10.1056/NEJMsr2214184}}.
\bibitem[{Li et~al.(2025)Li, Castulo and Xu}]{li2025embracing}
\bibinfo{author}{Li, Y.}, \bibinfo{author}{Castulo, N.J.}, \bibinfo{author}{Xu,
  X.}, \bibinfo{year}{2025}.
\newblock \bibinfo{title}{Embracing or rejecting {AI? A} mixed-method study on
  undergraduate students’ perceptions of artificial intelligence at a private
  university in {China}}, in: \bibinfo{booktitle}{{Frontiers in Education}},
  \bibinfo{organization}{Frontiers Media SA}. p. \bibinfo{pages}{1505856}.
\newblock \URLprefix
  \url{https://www.frontiersin.org/journals/education/articles/10.3389/feduc.2025.1505856},
  \DOIprefix\doi{10.3389/feduc.2025.1505856}.
\bibitem[{Manikonda et~al.(2018)Manikonda, Deotale and
  Kambhampati}]{manikonda2018s}
\bibinfo{author}{Manikonda, L.}, \bibinfo{author}{Deotale, A.},
  \bibinfo{author}{Kambhampati, S.}, \bibinfo{year}{2018}.
\newblock \bibinfo{title}{What's up with privacy? {User} preferences and
  privacy concerns in intelligent personal assistants}, in:
  \bibinfo{booktitle}{{Proceedings of the 2018 AAAI/ACM Conference on AI,
  Ethics, and Society}}, pp. \bibinfo{pages}{229--235}.
\newblock \URLprefix \url{https://dl.acm.org/doi/abs/10.1145/3278721.3278773},
  \DOIprefix\doi{10.1145/3278721.3278773}.
\bibitem[{Masa'deh et~al.(2024)Masa'deh, Majali, Alkhaffaf, Thurasamy,
  Almajali, Altarawneh, Al-Sherideh and Altarawni}]{masa2024antecedents}
\bibinfo{author}{Masa'deh, R.}, \bibinfo{author}{Majali, S.A.},
  \bibinfo{author}{Alkhaffaf, M.}, \bibinfo{author}{Thurasamy, R.},
  \bibinfo{author}{Almajali, D.}, \bibinfo{author}{Altarawneh, K.},
  \bibinfo{author}{Al-Sherideh, A.}, \bibinfo{author}{Altarawni, I.},
  \bibinfo{year}{2024}.
\newblock \bibinfo{title}{Antecedents of adoption and usage of {ChatGPT among
  Jordanian} university students: Empirical study}.
\newblock \bibinfo{journal}{International Journal of Data and Network Science}
  \bibinfo{volume}{8}, \bibinfo{pages}{1099--1110}.
\newblock \URLprefix
  \url{https://www.growingscience.com/ijds/Vol8/ijdns_2023_224.pdf},
  \DOIprefix\doi{10.5267/j.ijdns.2023.11.024}.
\bibitem[{Mittelstadt(2019)}]{mittelstadt2019principles}
\bibinfo{author}{Mittelstadt, B.}, \bibinfo{year}{2019}.
\newblock \bibinfo{title}{Principles alone cannot guarantee ethical {AI}}.
\newblock \bibinfo{journal}{Nature Machine Intelligence} \bibinfo{volume}{1},
  \bibinfo{pages}{501--507}.
\newblock \URLprefix \url{https://www.nature.com/articles/s42256-019-0114-4},
  \DOIprefix\doi{10.1038/s42256-019-0114-4}.
\bibitem[{Naud{\'e}(2020)}]{naude2020artificial}
\bibinfo{author}{Naud{\'e}, W.}, \bibinfo{year}{2020}.
\newblock \bibinfo{title}{Artificial intelligence vs {COVID-19}: Limitations,
  constraints and pitfalls}.
\newblock \bibinfo{journal}{AI \& Society} \bibinfo{volume}{35},
  \bibinfo{pages}{761--765}.
\newblock \URLprefix
  \url{https://link.springer.com/article/10.1007/S00146-020-00978-0},
  \DOIprefix\doi{10.1007/S00146-020-00978-0}.
\bibitem[{Oc et~al.(2024)Oc, Gonsalves and Quamina}]{oc2024generative}
\bibinfo{author}{Oc, Y.}, \bibinfo{author}{Gonsalves, C.},
  \bibinfo{author}{Quamina, L.T.}, \bibinfo{year}{2024}.
\newblock \bibinfo{title}{Generative {AI} in higher education assessments:
  Examining risk and tech-savviness on student’s adoption}.
\newblock \bibinfo{journal}{Journal of Marketing Education} ,
  \bibinfo{pages}{1--18}\URLprefix
  \url{https://journals.sagepub.com/doi/10.1177/02734753241302459},
  \DOIprefix\doi{10.1177/02734753241302459}.
\bibitem[{Park and Park(2024)}]{park2024filter}
\bibinfo{author}{Park, H.W.}, \bibinfo{author}{Park, S.}, \bibinfo{year}{2024}.
\newblock \bibinfo{title}{The filter bubble generated by artificial
  intelligence algorithms and the network dynamics of collective polarization
  on {YouTube: The} case of {South Korea}}.
\newblock \bibinfo{journal}{Asian Journal of Communication}
  \bibinfo{volume}{34}, \bibinfo{pages}{195--212}.
\newblock \URLprefix \url{https://doi.org/10.1080/01292986.2024.2315584},
  \DOIprefix\doi{10.1080/01292986.2024.2315584}.
\bibitem[{Pierson(2017)}]{pierson2017demographics}
\bibinfo{author}{Pierson, E.}, \bibinfo{year}{2017}.
\newblock \bibinfo{title}{Demographics and discussion influence views on
  algorithmic fairness}.
\newblock \bibinfo{journal}{arXiv preprint arXiv:1712.09124} \URLprefix
  \url{https://doi.org/10.48550/arXiv.1712.09124},
  \DOIprefix\doi{10.48550/arXiv.1712.09124}.
\bibitem[{Raghavan et~al.(2020)Raghavan, Barocas, Kleinberg and
  Levy}]{raghavan2020mitigating}
\bibinfo{author}{Raghavan, M.}, \bibinfo{author}{Barocas, S.},
  \bibinfo{author}{Kleinberg, J.}, \bibinfo{author}{Levy, K.},
  \bibinfo{year}{2020}.
\newblock \bibinfo{title}{Mitigating bias in algorithmic hiring: Evaluating
  claims and practices}, in: \bibinfo{booktitle}{{Proceedings of the 2020
  Conference on Fairness, Accountability, and Transparency}}, pp.
  \bibinfo{pages}{469--481}.
\newblock \URLprefix \url{https://dl.acm.org/doi/abs/10.1145/3351095.3372828},
  \DOIprefix\doi{10.1145/3351095.3372828}.
\bibitem[{Rawashdeh(2025)}]{rawashdeh2025consequences}
\bibinfo{author}{Rawashdeh, A.}, \bibinfo{year}{2025}.
\newblock \bibinfo{title}{The consequences of artificial intelligence: An
  investigation into the impact of {AI} on job displacement in accounting}.
\newblock \bibinfo{journal}{Journal of Science and Technology Policy
  Management} \bibinfo{volume}{16}, \bibinfo{pages}{506--535}.
\newblock \URLprefix \url{https://doi.org/10.1108/JSTPM-02-2023-0030},
  \DOIprefix\doi{10.1108/JSTPM-02-2023-0030},
  \href{http://arxiv.org/abs/https://www.emerald.com/jstpm/article-pdf/16/3/506/9659776/jstpm-02-2023-0030.pdf}{{\tt
  arXiv:https://www.emerald.com/jstpm/article-pdf/16/3/506/9659776/jstpm-02-2023-0030.pdf}}.
\bibitem[{Roselli et~al.(2019)Roselli, Matthews and
  Talagala}]{roselli2019managing}
\bibinfo{author}{Roselli, D.}, \bibinfo{author}{Matthews, J.},
  \bibinfo{author}{Talagala, N.}, \bibinfo{year}{2019}.
\newblock \bibinfo{title}{Managing bias in {AI}}, in:
  \bibinfo{booktitle}{{Companion Proceedings of the 2019 World Wide Web
  Conference}}, pp. \bibinfo{pages}{539--544}.
\newblock \URLprefix \url{https://dl.acm.org/doi/abs/10.1145/3308560.3317590},
  \DOIprefix\doi{10.1145/3308560.3317590}.
\bibitem[{S{\'a}ez-Velasco et~al.(2025)S{\'a}ez-Velasco,
  Alaguero-Rodr{\'\i}guez, Rodr{\'\i}guez-Cano and
  Delgado-Benito}]{saez2025students}
\bibinfo{author}{S{\'a}ez-Velasco, S.},
  \bibinfo{author}{Alaguero-Rodr{\'\i}guez, M.},
  \bibinfo{author}{Rodr{\'\i}guez-Cano, S.}, \bibinfo{author}{Delgado-Benito,
  V.}, \bibinfo{year}{2025}.
\newblock \bibinfo{title}{Students' attitudes towards {AI} and how they
  perceive the effectiveness of {AI} in designing video games}.
\newblock \bibinfo{journal}{Sustainability} \bibinfo{volume}{17},
  \bibinfo{pages}{3096}.
\newblock \URLprefix \url{https://www.mdpi.com/2071-1050/17/7/3096},
  \DOIprefix\doi{10.3390/su17073096}.
\bibitem[{Saheb(2023)}]{saheb2023ethically}
\bibinfo{author}{Saheb, T.}, \bibinfo{year}{2023}.
\newblock \bibinfo{title}{Ethically contentious aspects of artificial
  intelligence surveillance: A social science perspective}.
\newblock \bibinfo{journal}{AI and Ethics} \bibinfo{volume}{3},
  \bibinfo{pages}{369--379}.
\newblock \URLprefix
  \url{https://link.springer.com/article/10.1007/s43681-022-00196-y},
  \DOIprefix\doi{10.1007/s43681-022-00196-y}.
\bibitem[{Said et~al.(2023)Said, Potinteu, Brich, Buder, Schumm and
  Huff}]{said2023artificial}
\bibinfo{author}{Said, N.}, \bibinfo{author}{Potinteu, A.E.},
  \bibinfo{author}{Brich, I.}, \bibinfo{author}{Buder, J.},
  \bibinfo{author}{Schumm, H.}, \bibinfo{author}{Huff, M.},
  \bibinfo{year}{2023}.
\newblock \bibinfo{title}{An artificial intelligence perspective: How knowledge
  and confidence shape risk and benefit perception}.
\newblock \bibinfo{journal}{Computers in Human Behavior} \bibinfo{volume}{149},
  \bibinfo{pages}{107855}.
\newblock \URLprefix \url{https://doi.org/10.1016/j.chb.2023.107855},
  \DOIprefix\doi{10.1016/j.chb.2023.107855}.
\bibitem[{Schei et~al.(2024)Schei, M{\o}gelvang and
  Ludvigsen}]{schei2024perceptions}
\bibinfo{author}{Schei, O.M.}, \bibinfo{author}{M{\o}gelvang, A.},
  \bibinfo{author}{Ludvigsen, K.}, \bibinfo{year}{2024}.
\newblock \bibinfo{title}{Perceptions and use of {AI} chatbots among students
  in higher education: A scoping review of empirical studies}.
\newblock \bibinfo{journal}{Education Sciences} \bibinfo{volume}{14},
  \bibinfo{pages}{922}.
\newblock \URLprefix \url{https://www.mdpi.com/2227-7102/14/8/922},
  \DOIprefix\doi{10.3390/educsci14080922}.
\bibitem[{Shaheen(2021)}]{shaheen2021applications}
\bibinfo{author}{Shaheen, M.Y.}, \bibinfo{year}{2021}.
\newblock \bibinfo{title}{Applications of artificial intelligence {(AI)} in
  healthcare: A review}.
\newblock \bibinfo{howpublished}{ScienceOpen Preprints}.
\newblock \bibinfo{note}{Preprint.
  \url{https://www.scienceopen.com/document?vid=b9349a6e-e2f0-4cf3-ae4e-efb4cda01690}}.
\bibitem[{Slovic(2016)}]{slovic2016perception}
\bibinfo{author}{Slovic, P.}, \bibinfo{year}{2016}.
\newblock \bibinfo{title}{Perception of risk}, in: \bibinfo{booktitle}{The
  perception of risk}. \bibinfo{publisher}{Routledge}, pp.
  \bibinfo{pages}{220--231}.
\newblock \URLprefix
  \url{https://www.taylorfrancis.com/chapters/edit/10.4324/9781315661773-13/perception-risk-paul-slovic}.
\bibitem[{Stahl and Stahl(2021)}]{stahl2021ethical}
\bibinfo{author}{Stahl, B.C.}, \bibinfo{author}{Stahl, B.C.},
  \bibinfo{year}{2021}.
\newblock \bibinfo{title}{Ethical issues of {AI}}.
\newblock \bibinfo{journal}{Artificial Intelligence for a Better Future: An
  ecosystem Perspective on the Ethics of AI and Emerging Digital Technologies}
  , \bibinfo{pages}{35--53}\URLprefix
  \url{https://link.springer.com/chapter/10.1007/978-3-030-69978-9_4},
  \DOIprefix\doi{10.1007/978-3-030-69978-9_4}.
\bibitem[{Stöhr et~al.(2024)Stöhr, Ou and Malmström}]{STOHR2024100259}
\bibinfo{author}{Stöhr, C.}, \bibinfo{author}{Ou, A.W.},
  \bibinfo{author}{Malmström, H.}, \bibinfo{year}{2024}.
\newblock \bibinfo{title}{Perceptions and usage of ai chatbots among students
  in higher education across genders, academic levels and fields of study}.
\newblock \bibinfo{journal}{Computers and Education: Artificial Intelligence}
  \bibinfo{volume}{7}, \bibinfo{pages}{100259}.
\newblock \URLprefix
  \url{https://www.sciencedirect.com/science/article/pii/S2666920X24000626},
  \DOIprefix\doi{10.1016/j.caeai.2024.100259}.
\bibitem[{Tierney et~al.(2025)Tierney, Peasey and Gould}]{tierney2025student}
\bibinfo{author}{Tierney, A.}, \bibinfo{author}{Peasey, P.},
  \bibinfo{author}{Gould, J.}, \bibinfo{year}{2025}.
\newblock \bibinfo{title}{Student perceptions on the impact of {AI} on their
  teaching and learning experiences in higher education}.
\newblock \bibinfo{journal}{Research \& Practice in Technology Enhanced
  Learning} \bibinfo{volume}{20}, \bibinfo{pages}{1--25}.
\newblock \DOIprefix\doi{10.58459/rptel.2025.20005}.
\bibitem[{Torres et~al.(2024)Torres, Wenke, Lieneck, Ramamonjiarivelo and
  Ari}]{torres2024systematic}
\bibinfo{author}{Torres, A.}, \bibinfo{author}{Wenke, M.},
  \bibinfo{author}{Lieneck, C.}, \bibinfo{author}{Ramamonjiarivelo, Z.},
  \bibinfo{author}{Ari, A.}, \bibinfo{year}{2024}.
\newblock \bibinfo{title}{A systematic review of artificial intelligence used
  to predict loneliness, social isolation, and drug use during the {COVID-19}
  pandemic}.
\newblock \bibinfo{journal}{Journal of Multidisciplinary Healthcare} ,
  \bibinfo{pages}{3403--3425}\URLprefix
  \url{https://www.tandfonline.com/doi/full/10.2147/JMDH.S466099},
  \DOIprefix\doi{10.2147/JMDH.S466099}.
\bibitem[{Trope and Liberman(2010)}]{trope2010construal}
\bibinfo{author}{Trope, Y.}, \bibinfo{author}{Liberman, N.},
  \bibinfo{year}{2010}.
\newblock \bibinfo{title}{Construal-level theory of psychological distance}.
\newblock \bibinfo{journal}{Psychological review} \bibinfo{volume}{117},
  \bibinfo{pages}{440}.
\newblock \URLprefix
  \url{https://psycnet.apa.org/fulltext/2010-06891-005.html},
  \DOIprefix\doi{10.1037/a0018963}.
\bibitem[{{UNESCO}(2021)}]{UNESCOAI2021}
\bibinfo{author}{{UNESCO}}, \bibinfo{year}{2021}.
\newblock \bibinfo{title}{Recommendation on the ethics of artificial
  intelligence}.
\newblock
  \bibinfo{howpublished}{\url{https://unesdoc.unesco.org/ark:/48223/pf0000381137}}.
\bibitem[{Venkatesh et~al.(2003)Venkatesh, Morris, Davis and
  Davis}]{venkatesh2003user}
\bibinfo{author}{Venkatesh, V.}, \bibinfo{author}{Morris, M.G.},
  \bibinfo{author}{Davis, G.B.}, \bibinfo{author}{Davis, F.D.},
  \bibinfo{year}{2003}.
\newblock \bibinfo{title}{User acceptance of information technology: Toward a
  unified view}.
\newblock \bibinfo{journal}{MIS quarterly} ,
  \bibinfo{pages}{425--478}\URLprefix
  \url{https://misq.umn.edu/misq/article-abstract/27/3/425/1340/User-Acceptance-of-Information-Technology-Toward-A},
  \DOIprefix\doi{10.2307/30036540}.
\bibitem[{Von~Eschenbach(2021)}]{von2021transparency}
\bibinfo{author}{Von~Eschenbach, W.J.}, \bibinfo{year}{2021}.
\newblock \bibinfo{title}{Transparency and the black box problem: Why we do not
  trust {AI}}.
\newblock \bibinfo{journal}{Philosophy \& Technology} \bibinfo{volume}{34},
  \bibinfo{pages}{1607--1622}.
\newblock \URLprefix
  \url{https://link.springer.com/article/10.1007/s13347-021-00477-0},
  \DOIprefix\doi{10.1007/s13347-021-00477-0}.
\bibitem[{Wang et~al.(2021)Wang, Liu and Tu}]{wang2021factors}
\bibinfo{author}{Wang, Y.}, \bibinfo{author}{Liu, C.}, \bibinfo{author}{Tu,
  Y.F.}, \bibinfo{year}{2021}.
\newblock \bibinfo{title}{Factors affecting the adoption of ai-based
  applications in higher education}.
\newblock \bibinfo{journal}{Educational technology \& society}
  \bibinfo{volume}{24}, \bibinfo{pages}{116--129}.
\newblock \URLprefix \url{https://www.jstor.org/stable/27032860}.
\bibitem[{Yu et~al.(2023)Yu, Bentley and Carroll}]{yu2023enhancing}
\bibinfo{author}{Yu, S.}, \bibinfo{author}{Bentley, B.L.},
  \bibinfo{author}{Carroll, F.}, \bibinfo{year}{2023}.
\newblock \bibinfo{title}{Enhancing smart home security: A privacy risk
  analysis framework}, in: \bibinfo{booktitle}{{International Conference on
  Cyber Security, Privacy in Communication Networks}},
  \bibinfo{organization}{Springer}. pp. \bibinfo{pages}{295--308}.
\newblock \URLprefix \url{https://doi.org/10.1007/978-981-97-3973-8_18},
  \DOIprefix\doi{10.1007/978-981-97-3973-8_18}.
\bibitem[{Yu and Carroll(2022a)}]{yu2022implications}
\bibinfo{author}{Yu, S.}, \bibinfo{author}{Carroll, F.}, \bibinfo{year}{2022}a.
\newblock \bibinfo{title}{Implications of {AI} in national security:
  understanding the security issues and ethical challenges}, in:
  \bibinfo{booktitle}{{Artificial Intelligence in Cyber Security: Impact and
  Implications: Security Challenges, Technical and Ethical Issues, Forensic
  Investigative Challenges}}. \bibinfo{publisher}{Springer}, pp.
  \bibinfo{pages}{157--175}.
\newblock \URLprefix \url{https://doi.org/10.1007/978-3-030-88040-8_6},
  \DOIprefix\doi{10.1007/978-3-030-88040-8_6}.
\bibitem[{Yu and Carroll(2022b)}]{yu2022insights}
\bibinfo{author}{Yu, S.}, \bibinfo{author}{Carroll, F.}, \bibinfo{year}{2022}b.
\newblock \bibinfo{title}{Insights into the next generation of policing:
  understanding the impact of technology on the police force in the digital
  age}, in: \bibinfo{booktitle}{{Artificial Intelligence and National
  Security}}. \bibinfo{publisher}{Springer}, pp. \bibinfo{pages}{169--191}.
\newblock \URLprefix \url{https://doi.org/10.1007/978-3-031-06709-9_9},
  \DOIprefix\doi{10.1007/978-3-031-06709-9_9}.
\bibitem[{Yu and Carroll(2023)}]{yu2023balance}
\bibinfo{author}{Yu, S.}, \bibinfo{author}{Carroll, F.}, \bibinfo{year}{2023}.
\newblock \bibinfo{title}{A balance of power: Exploring the opportunities and
  challenges of {AI} for a nation}, in: \bibinfo{booktitle}{Applications for
  Artificial Intelligence and Digital Forensics in National Security}.
  \bibinfo{publisher}{Springer}, pp. \bibinfo{pages}{15--37}.
\newblock \URLprefix \url{https://doi.org/10.1007/978-3-031-40118-3_2},
  \DOIprefix\doi{10.1007/978-3-031-40118-3_2}.
\bibitem[{Yu et~al.(2024a)Yu, Carroll and Bentley}]{yu2024insights}
\bibinfo{author}{Yu, S.}, \bibinfo{author}{Carroll, F.},
  \bibinfo{author}{Bentley, B.L.}, \bibinfo{year}{2024}a.
\newblock \bibinfo{title}{Insights into privacy protection research in {AI}}.
\newblock \bibinfo{journal}{IEEE Access} \bibinfo{volume}{12},
  \bibinfo{pages}{41704--41726}.
\newblock \URLprefix \url{https://doi.org/10.1109/ACCESS.2024.3378126},
  \DOIprefix\doi{10.1109/ACCESS.2024.3378126}.
\bibitem[{Yu et~al.(2024b)Yu, Carroll and Bentley}]{yu2024trust}
\bibinfo{author}{Yu, S.}, \bibinfo{author}{Carroll, F.},
  \bibinfo{author}{Bentley, B.L.}, \bibinfo{year}{2024}b.
\newblock \bibinfo{title}{Trust and risk: Psybersecurity in the {AI} era}, in:
  \bibinfo{booktitle}{{Psybersecurity}}. \bibinfo{publisher}{CRC Press}, pp.
  \bibinfo{pages}{130--155}.
\newblock \URLprefix \url{https://doi.org/10.1201/9781032664859-6},
  \DOIprefix\doi{10.1201/9781032664859-6}.
\bibitem[{Yu et~al.(2024c)Yu, Carroll and Bentley}]{yu2024chatgpt}
\bibinfo{author}{Yu, S.}, \bibinfo{author}{Carroll, F.},
  \bibinfo{author}{Bentley, B.L.}, \bibinfo{year}{2024}c.
\newblock \bibinfo{title}{Trust and trustworthiness: Privacy protection in the
  {ChatGPT} era}, in: \bibinfo{booktitle}{Data Protection: The Wake of AI and
  Machine Learning}. \bibinfo{publisher}{Springer}, pp.
  \bibinfo{pages}{103--127}.
\newblock \URLprefix \url{https://doi.org/10.1007/978-3-031-76473-8_6},
  \DOIprefix\doi{10.1007/978-3-031-76473-8_6}.
\bibitem[{Zajko(2022)}]{zajko2022artificial}
\bibinfo{author}{Zajko, M.}, \bibinfo{year}{2022}.
\newblock \bibinfo{title}{Artificial intelligence, algorithms, and social
  inequality: Sociological contributions to contemporary debates}.
\newblock \bibinfo{journal}{Sociology Compass} \bibinfo{volume}{16},
  \bibinfo{pages}{e12962}.
\newblock \URLprefix
  \url{https://compass.onlinelibrary.wiley.com/doi/full/10.1111/soc4.12962},
  \DOIprefix\doi{10.1111/soc4.12962}.
\bibitem[{Zhai et~al.(2024)Zhai, Wibowo and Li}]{zhai2024effects}
\bibinfo{author}{Zhai, C.}, \bibinfo{author}{Wibowo, S.}, \bibinfo{author}{Li,
  L.D.}, \bibinfo{year}{2024}.
\newblock \bibinfo{title}{The effects of over-reliance on {AI} dialogue systems
  on students' cognitive abilities: A systematic review}.
\newblock \bibinfo{journal}{Smart Learning Environments} \bibinfo{volume}{11},
  \bibinfo{pages}{28}.
\newblock \URLprefix
  \url{https://link.springer.com/article/10.1186/s40561-024-00316-7},
  \DOIprefix\doi{10.1186/s40561-024-00316-7}.
\bibitem[{Zhai et~al.(2021)Zhai, Chu, Chai, Jong, Istenic, Spector, Liu, Yuan
  and Li}]{zhai2021review}
\bibinfo{author}{Zhai, X.}, \bibinfo{author}{Chu, X.}, \bibinfo{author}{Chai,
  C.S.}, \bibinfo{author}{Jong, M.S.Y.}, \bibinfo{author}{Istenic, A.},
  \bibinfo{author}{Spector, M.}, \bibinfo{author}{Liu, J.B.},
  \bibinfo{author}{Yuan, J.}, \bibinfo{author}{Li, Y.}, \bibinfo{year}{2021}.
\newblock \bibinfo{title}{A review of artificial intelligence {(AI)} in
  education from 2010 to 2020}.
\newblock \bibinfo{journal}{Complexity} \bibinfo{volume}{2021},
  \bibinfo{pages}{1--18}.
\newblock \URLprefix
  \url{https://onlinelibrary.wiley.com/doi/full/10.1155/2021/8812542},
  \DOIprefix\doi{10.1155/2021/8812542}.

\end{thebibliography}

\end{document}